\documentclass[]{cupconf}
\usepackage{graphicx}

\title[Massive stars vs. nebular abundances in the Orion nebula]
      {Massive stars vs. nebular abundances\\ in the Orion nebula}
      
\author[S.~Sim\'on-D\'iaz]{\\ S. Sim\'on-D\'iaz$^{1, 2}$ 
} 
\affiliation{$^1$ Instituto de Astrof\'{\i}sica de Canarias, E-38200 La Laguna, Tenerife, Spain \\ 
             $^2$ LUTH, Observatoire de Meudon, 92195 Meudon Cedex, France \\ }   

\begin{document}
\maketitle

\begin{abstract}
The search of consistency between nebular and massive star abundances has 
been a longstanding problem. I briefly review what has been 
done regarding to this topic, also presenting a recent study focused on the
Orion nebula: the O and Si stellar abundances resulting from a detailed 
and fully consistent spectroscopic analysis of the group of B stars 
associated with the Orion nebula are compared with the most recent nebular 
gas-phase results.
\end{abstract}

\firstsection
\section{Introduction}

Photospheres of OB stars are representative of the interstellar 
material from which they were born due to their relative youth. The evolutionary 
characteristics of blue massive stars imply that these objects and the associated 
ionized nebulae - {\sc H\,ii} regions - must share the same chemical composition 
\footnote{There are certain cases in which this in not completely fulfilled:
(1) Strong stellar mass-loss may expose underlaying layers already contaminated 
on the surface of the star; (2) Several studies of OBA-type stars have found observational evidences of stellar surface contamination
by products from the CNO-bicycle; rotating models by Maeder \& 
Meynet (2000) predict that mixing of nuclear processed material at the
surface will increase with stellar mass, age, initial rotational velocity and
decreasing metallicity; (3) Certain elements in the nebular material can be 
depleted onto dust grains; in this case
the gas-phase abundances, derived through a spectroscopic study of the 
{\sc H\,ii} region, could be somewhat lower than the stellar ones.} 

Traditionally, chemical abundance studies in spiral and irregular galaxies have been based on the emission line spectra 
of {\sc H\,ii} regions. This is logical, since {\sc H\,ii} regions are 
luminous and have high surface brightness (in the emission lines) compared
to individual stars in galaxies. 
Therefore, it is relatively easy to obtain high
quality spectroscopic observational data, even with small and medium
size telescopes. This has made it possible both the detailed study of 
individual nebulae (viz. Esteban et al. 2004), and the determination of
radial gradients in the Milky way (viz. Shaver et al. 1983, Afflerbach
et al. 1997, Esteban et al. 2005) and other spiral galaxies (see the 
invited talk by F. Bresolin in these proceedings), imposing observational constraints
to the chemical evolution models of these galaxies (see review
by L. Carigi in these proceedings).

Although this is a commonly used methodology, it is not without some 
difficulties and problems. For example, it is known that the optical 
recombination lines ({\sc orl}) in ionized nebulae indicate higher
abundances than collisionally excited lines ({\sc cel}); temperature 
fluctuations, density condensations and abundance inhomogeneities
have been proposed to solve the {\sc orl/cel} problem, however 
none of these explanations are completely satisfactory (see Esteban 2002
for a review). I refer to F. Bresolin talk (this proceedings) for a
review of the use and limitations of strong line methods in the 
determination of nebular abundances. Stasi\'nska (2005) has recently shown 
that for metal rich nebulae, the derived abundances based on a direct measurements
of the electron temperature (T$_{\rm e}$) may deviate systematically from the real ones. Finally, 
one must keep in mind other sources of uncertainties in the nebular 
abundance determination such as the atomic data, reddening corrections 
and the possible depletion of elements onto dust.

Among the stellar objects, blue massive stars can be easily identified 
at large distances due to their high luminosities. 
Therefore, massive stars offer a unique opportunity to study 
present-day chemical abundances in spiral and irregular galaxies as an alternative 
method to classical {\sc H\,ii} region studies.
However, the amount of energy released by these stars is so large that it 
produces dramatic effects on the star itself: these stellar objects show
mass outflows during their whole lifetime (so called stellar winds) and
their atmospheres departure from {\sc lte} conditions, two facts that make
their modeling quite complex.
It has not been until very recently that the development of massive star
model atmospheres and the growth of the computational efficiency has
allowed a reliable abundance analysis of these objects. Nowadays, the 
later is feasible; however, one must take into account that there are some 
effects that can affect the final results and must be carefully treated: 
(1) hypothesis on the stellar atmosphere modeling ({\sc lte} vs. {\sc nlte}, 
plane-parallel vs. spherical models, inclusion of line-blanketing effects), 
(2) atomic models and atomic data; (3) establishment of the stellar parameters 
and microturbulence.

Both nebular and stellar methodologies are now working in tandem 
to progress in our knowledge of the 
metallicity content of irregular and spiral galaxies (from the Milky Way 
to far beyond the Local Group). Since they sample similar spatial and temporal 
distributions, these objects offers us a unique framework in which the 
reliability of the derived abundances in both methodologies can be tested.

\section{Do stellar and nebular abundances agree?}

Two type of studies can be addressed to investigate if there is a consistency
of abundance results from stellar and nebular studies: the 
comparison of radial abundance gradients in spiral galaxies or mean abundances
in irregular galaxies ({\em global approach}), and the comparison of absolute
abundances in the same galactic region ({\em local approach}).
Values found in the literature for the oxygen abundance gradient in the Galaxy
range from -0.04 to -0.08 in the nebular case (Esteban et al. 2005, and 
references therein), and
from -0.03 to -0.07 in the stellar case (Gummersbach et al. 1998, Rolleston
et al. 2000, Daflon \& Cunha 2004). One could conclude that results from both 
methodologies are in agreement within the intrinsic uncertainties, however
this may not be the case when comparing results from individual studies
(see e.g Figure 1). 

In the extragalactic case, Urbaneja et al. (2005b) found good agreement in
the comparison of stellar and {\sc Hii} region O abundances (based on 
direct determinations of the T$_{\rm e}$ of the nebulae) in M33; 
however, other works in spiral galaxies has found that the stellar and nebular 
agreement tend to be very dependent on the calibrations used in the strong 
nebular methods (Trundle et al. 2002, Urbaneja et al. 2005a).

Examples of the comparison between stellar and nebular abundances in the same 
galactic region or mean abundances in the Magellanic Clouds are found in
Cunha \& Lambert (1994), Korn et al. (2002) and Trundle et al. (2004). 
Although within the typical errors of the stellar and nebular analyses
the abundances determined by both methodologies are fairly consistent, more work
remains necessary. For example, until very recently, there has not been any
consistent detailed comparison of stellar and nebular abundances in the same
star forming region. 

Within this aim Sim\'on-D\'iaz et al (2006) selected the Orion nebula, 
a well studied and spatially resolved {\sc H\,ii} region with a cluster 
of a few massive stars inside it (the Trapezium cluster). Below, I will 
present the results of a detailed and fully consistent 
spectroscopic analysis of three B0.5V stars associated with
the Orion nebula for deriving their O and Si abundances. The resulting 
abundances are compared with the most recent nebular gas-phase results.

\begin{figure}
\center
\includegraphics[height=2.1in]{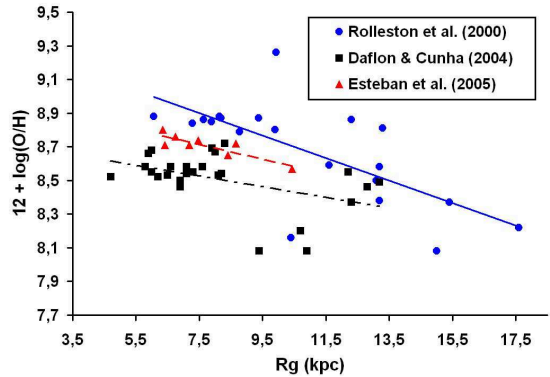}
\caption{Comparison of three recent determinations of the O abundance gradient in the
Milky Way. Although the comparison of values found in the literature for the
gradient obtained by means of massive stars and {\sc Hii} region studies seems 
to be in agreement within the intrinsic uncertainties (see text), this may 
not be the case when comparing results from individual studies. Note also the
offset in absolute abundances resulting from the three studies.}\label{fig}
\end{figure}

\section{Massive stars and nebular abundances in Orion nebula}

The stellar spectroscopic observational dataset consists of {\sc int@\,ids\,2.5}m
spectra of $\theta^1$\,Ori A, $\theta^1$\,Ori D, $\theta^2$\,Ori B, three
B0.5V stars located inside the Orion nebula.
Details on the spectroscopic analysis for the stellar parameters and oxygen 
abundance determination can be found in Sim\'on-D\'iaz et al (2006).
The analysis of silicon abundances were presented in Sim\'on-D\'iaz 
(2005, thesis), and will be published soon. Both stellar parameters and 
abundances where determined by means of the stellar atmosphere code 
{\sc fastwind} (Santolaya et al. 1997, Puls et al. 2005), a {\sc nlte} 
code taking into account line blanketing and stellar wind effects. 
A detailed O and Si abundance analysis by multiplets was previously performed,
using the slow rotator B0.2V star $\tau$\,Sco, to correctly estimate the abundance 
uncertainties related to the stellar parameters, microturbulence
and line-to-line abundance dispersion.

Table 1 summarizes the results from the stellar abundance determination, 
along with the values found in the literature for the nebular gas-phase 
abundances. The most recent and complete analysis of the chemical composition 
of the Orion nebula has been presented by Esteban et al. (2004); these authors 
used a wide variety of collisionally excited and recombination lines from 
a very deep {\sc uves@\,vlt\,8.2}m spectra for the abundance determination.
Table 1 shows the oxygen gas-phase abundance proposed by these authors, along 
with their estimated gas+dust oxygen abundance, which takes into account a dust 
deplection factor of 0.08 dex (obtained by Esteban et al. 1998).
For the nebular silicon abundance determination I refer to the works by
Rubin et al. (1993) and Garnett et al. (1995); these authors estimated the
Si abundances in the Orion nebula by using {\sc iue} high-dispersion spectra.

\section{Results and discussion in the context of the {\sc mru}}

The mean value of the stellar oxygen abundances derived by Sim\'on-D\'iaz 
et al. (2006) is in agreement with the nebular gas-phase oxygen abundance 
proposed by Esteban et al. (2004), however the dust+gas nebular oxygen 
abundance is somewhat larger than the stellar value. These result may imply a 
lower oxygen dust depletion factor than previously considered for the Orion 
nebula.

In the case of silicon abundances, the stellar results are systematically 
larger than the nebular values. This results suggests that a certain amount 
of nebular silicon is depleted onto dust grains; however, the exact 
determination of the silicon depletion factor is complicated due to 
uncertainties associated with the derived silicon abundances, both stellar 
and nebular.

I would like to finally remark the importance of this kind of comparisons for the study
of the Metal Rich Universe. In high metallicity environments, nebular T$_{\rm e}$ diagnostic lines 
are faint and hence hardly measurable, so strong line methods (which depend
on calibrations) must be used. Moreover, even 
if very deep spectra are obtained, derived nebular abundances may be affected 
by biases (Stasi\'nska 2005). These deficiencies in the nebular methodology can be
solved by using massive stars, since the strength of metal lines in the stellar spectra 
increases with metallicity, and hence stellar abundance diagnostic lines are 
(even!) more clearly seen in those high metallicity environments. 

\begin{table*}
\begin{center}
\begin{tabular}{c c l | r}
\hline
 Ref & Method & \ Abundance & Comments \\
\hline
(1) \ & \ \ Orion nebula B0.5V stars \ & \ $\epsilon$(O)\,=\,8.63\,$\pm$\,0.10 dex \ \ & $\theta^1$ Ori A, D, $\theta^2$ Ori B  \\ 
(2)   & Gas phase                      & \ $\epsilon$(O)\,=\,8.65\,$\pm$\,0.03 dex     & {\sc rl}, {\sc cel}, t$^2$\,=\,0.022   \\ 
(2)   & Gas+dust                       & \ $\epsilon$(O)\,=\,8.73\,$\pm$\,0.03 dex     & \ \ \ Dust depletion factor of 0.08    \\ 
\hline
(3)   & Orion nebula B0.5V stars       & \ $\epsilon$(Si)\,=\,7.55\,$\pm$\,0.20 dex    &  $\theta^1$ Ori A, D, $\theta^2$ Ori B \\ 
(4)   & Gas phase                      & \ $\epsilon$(Si)\,=\,6.65 dex                 &                                        \\ 
(5)   & Gas phase                      & \ $\epsilon$(Si)\,=\,6.60-7.15 dex            &                                        \\ 
\hline
\end{tabular}
\caption[]{Results from the stellar and nebular abundance analyses in the Orion 
           nebula:
           (1) Sim\'on-D\'iaz et al. (2006), 
		   (2) Esteban et al. (2004), 
		   (3) Sim\'on-D\'iaz (2005, thesis), 
		   (4) Rubin et al. (1993), 
		   (5) Garnett et al. (1995)}
\end{center}
\label{res}
\end{table*}

\end{document}